\documentclass[11pt,twoside]{article}
\usepackage{asp2006}
\usepackage{psfig}
\usepackage{epsf}
\usepackage{graphics}
\usepackage{lscape}
\def\edcomment#1{\iffalse\marginpar{\raggedright\sl#1\/}\else\relax\fi}
\reversemarginpar

\begin{document}
\title{SRVs in the Solar Neighbourhood}
\author{I.S. Glass}
\affil{South African Astronomical Observatory, PO Box 9, Observatory 7935,
South Africa}

\vspace{-2mm}

\begin{abstract}

Period-luminosity sequences have been shown to exist among the Semi-Regular
Variables (SRVs) of the Magellanic Clouds (Wood et al, 1999), the Bulge of
the Milky Way galaxy (Glass \& Schultheis, 2003) and elsewhere. It would
clearly be useful to have absolute (trigonometric) calibrations of these
relations. This paper investigates whether the sequences can be seen among
the M-type giant SRVs of the solar neighbourhood.  Mass loss phenomena among
these stars and their dependence on period and spectral sub-type are also
discussed.

\end{abstract}

The periods of 57 bright M-type SRVs have been extracted from Percy, Wilson
\& Henry (2001), Percy, Dunlop \& Kassim (2001), Percy, Nyssa \& Henry
(2001) and Percy et al (2004). These have been supplemented with information
from Bedding \& Zijlstra (1998) and CDS, Strasbourg. Parallaxes have been
taken from the Hipparcos Catalogue, via CDS, and $K$ magnitudes from
Neugebauer \& Leighton (1969). Fig 1 shows the resulting $M_K$, log$P$
diagram. Many stars have more than one period and all have been plotted. The
approximate positions of the sequences seen in the Bulge are given for
comparison.

The scatter in Fig 1 is rather large, but one can almost convince oneself
that the A and B sequences are separately visible. The D certainly is. The C
sequence, occupied by Miras, is not expected to be populated because of the
choice of sample; nevertheless a few objects do appear to lie on it. It is
hoped that more extensive photometric observations will in due course lead
to better magnitudes and periods and that improved parallax observations
will become available, so that the sequences will eventually be more
precisely defined.

IRAS 12$\mu$m magnitudes have been combined with those at $K$ to form $K$ --
[12] colours and these have been plotted against M sub-type in Fig 2 and
against log $P$ in Fig 3. As in the case of the NGC6522 field, infrared
excesses (due to mass-loss) appear in M-K types later than about M4 and at
periods longer than about 70 days (cf Alard et al, 2001, Glass \&
Schultheis, 2002).

I thank Dr T. Lloyd Evans, Dr M. Schultheis and Mr N. Matsunaga for useful 
comments during this work.

\vspace{-4mm}

 \setcounter{figure}{0}
 \begin{figure}[!h]
 \plotone{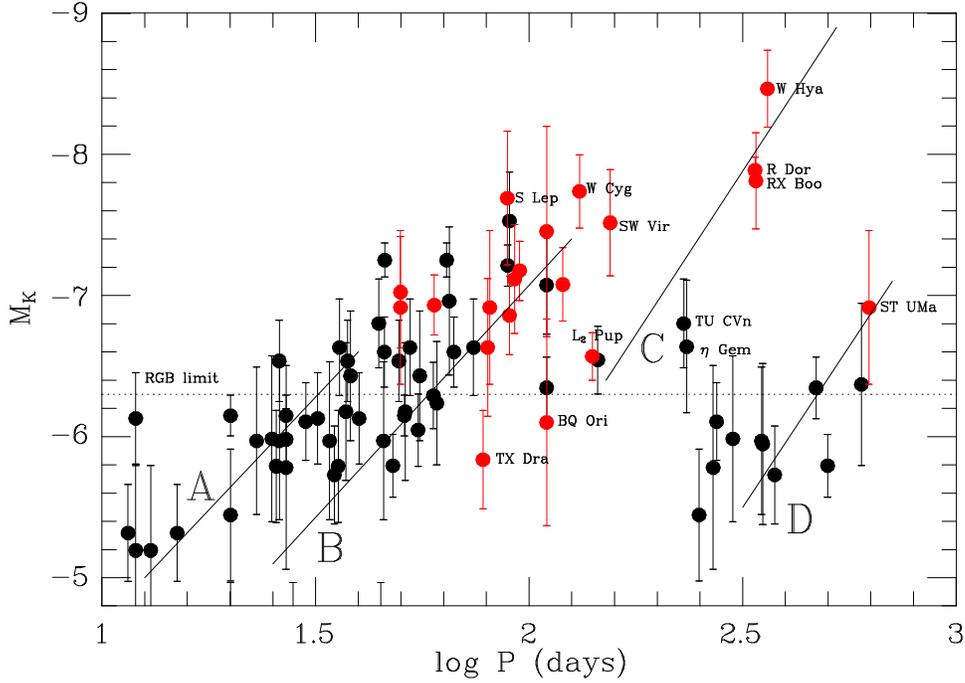}
 \caption{$M_K$, log$P$ diagram for SRVs in the solar neighbourhood. The
positions of the Galactic Bulge A,B,C and D sequences from Glass \&
Schultheis (2003) are shown. Stars plotted with light symbols have $K$ --
[12] $>$ 1.0, indicative of dust shells. Some interesting outliers have been
labelled.}
 \end{figure}

 \begin{figure}[!h]
 \plottwo{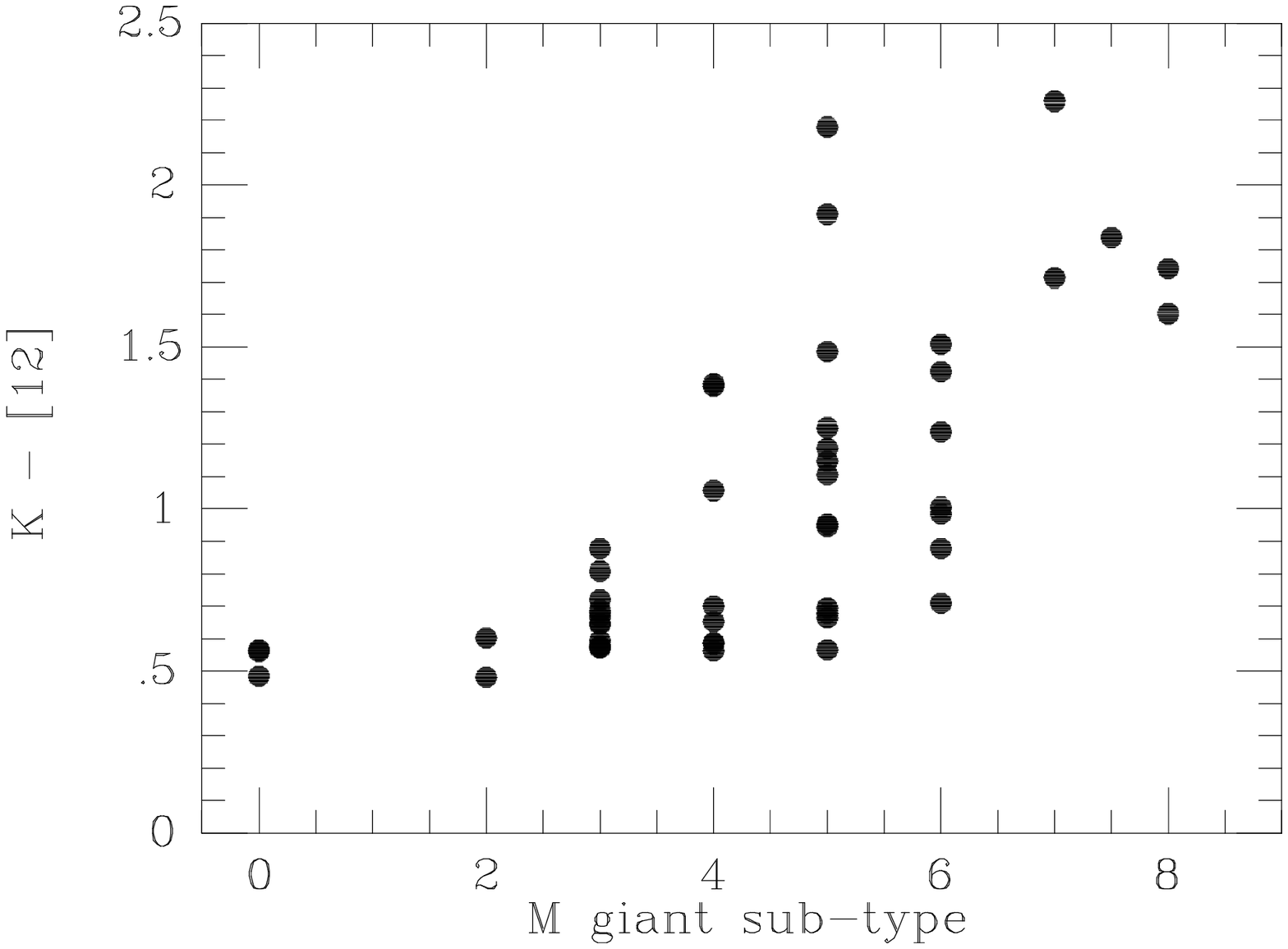}{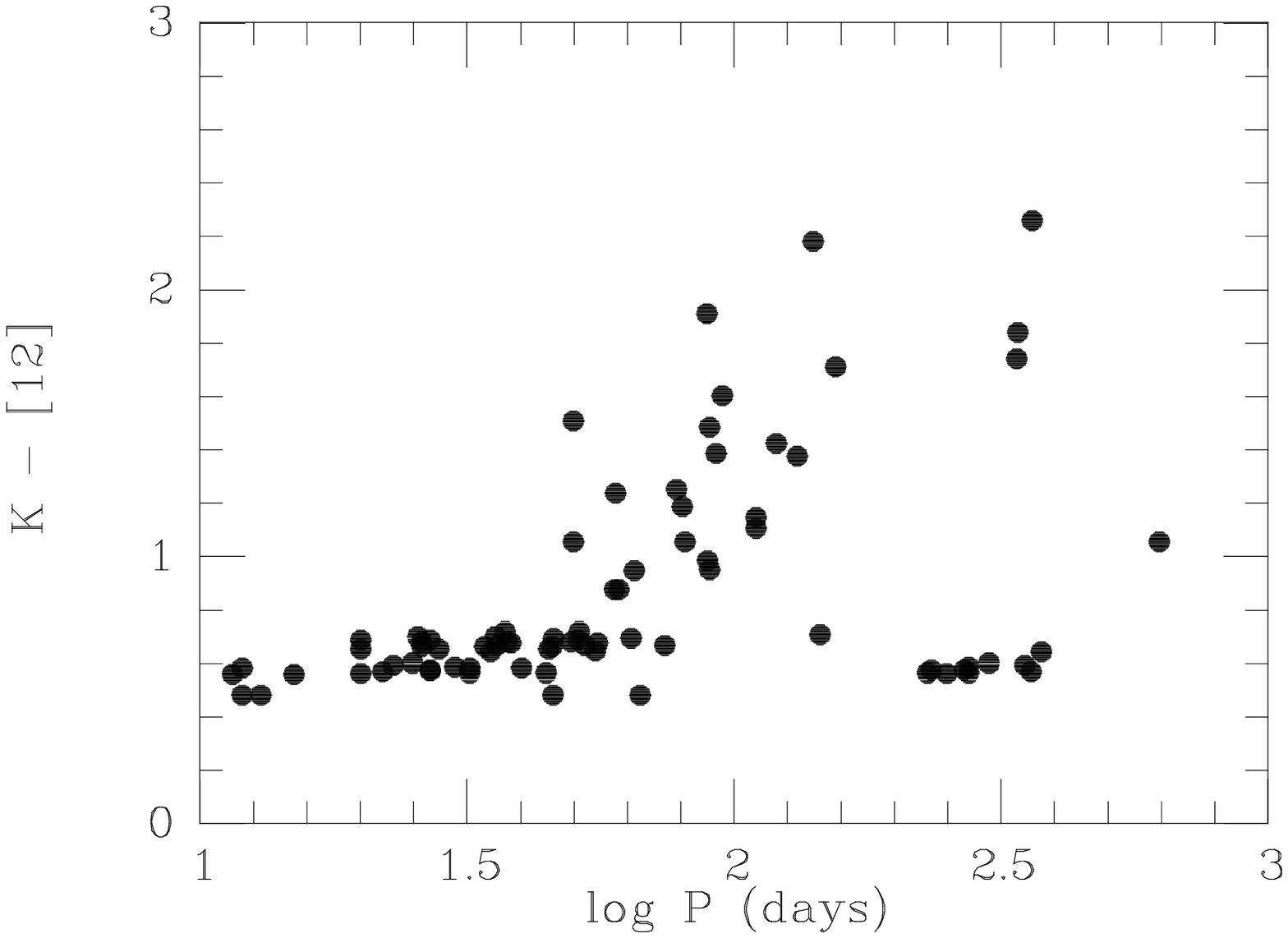}
 \caption{{\itshape Left:\/} $K$ -- [12] colour vs M-K type for the local
SRVs. Note the prevalence of significant excess 12 $\mu$m radiation beyond
types M3-M4.
 {\itshape Right:\/} $K$ -- [12] colour vs log period for local SRVs. All
periods are shown. The long-period D sequence stars usually do not exhibit
12 $\mu$m excess.} 
 \end{figure}


\begin{thebibliography}{}

\bibitem[]{}Alard, C. et al, 2001, ApJ, 552, 2001

\bibitem[]{}Bedding, T.R., Zijlstra, A.A. 1998, ApJ, 506, L47

\bibitem[]{}Glass, I.S., Schultheis, M. 2003, MNRAS, 345, 39

\bibitem[]{}Glass, I.S., Schultheis, M. 2002, MNRAS, 337, 519

\bibitem[]{}Neugebauer, G., Leighton, R.B. 1969, Two-Micron Sky Survey, NASA
SP-3047, Washington, D.C.

\bibitem[]{}Percy, J.R., Bakos, A.G., Besla, G., Hou, D., Velocci, V., Henry,
G.W. 2004, IAU Coll. 193, ASP Conf. Ser., 310, p348

\bibitem[]{}Percy, J.R., Nyssa, Z., Henry, G.W. 2001, IBVS 5209

\bibitem[]{}Percy, J.R., Wilson, J.B. and Henry, G.W. 2001, PASP, 113, 983

\bibitem[]{}Percy, J.R., Dunlop, H., Kassim, L., Thompson, R.R. 2001, IBVS
5041

\bibitem[]{}Wood, P.R. and the MACHO Team 1999, in IAU Symp.\ 191, p151 


\end{thebibliography}
\end{document}